\documentclass[12pt]{article}
\usepackage[colon]{natbib}  
\usepackage{authblk} 
\usepackage{graphicx}
\usepackage{epstopdf}
\usepackage{amssymb}
\usepackage{xspace}
\usepackage{fancyhdr}

\long\def\comment#1{} 
\newcommand{\ie}[0]{{\em i.e.}\xspace}
\newcommand{\eg}[0]{{\em e.g.}\xspace }

\newcommand{\vs}[0]{{\em vs.}\xspace}

\begin{document}
\title{Closed Form Jitter Analysis \\ of Neuronal Spike Trains}
\pagestyle{fancy}
\chead{Closed Form Jitter Analysis}
\rhead{}
\lhead{}
\author[1,2]{Daniel M. Jeck}

\author[1,3]{Ernst Niebur}
\affil[1]{Zanvyl Krieger Mind/Brain Institute, Johns Hopkins University, Baltimore, MD}
\affil[2]{Department of Biomedical Engineering, Johns
  Hopkins University, Baltimore, MD} 
\affil[3]{Solomon Snyder Department of Neuroscience, Johns Hopkins
  University, Baltimore, MD}

\maketitle

\newpage
{\bf Running Title} \\
Closed Form Jitter Analysis

{\bf Address correspondence to}\\
Daniel Jeck\\
The Johns Hopkins University\\
Krieger Mind/Brain Institute\\
3400 N. Charles Street\\
Baltimore, MD 21218\\
(email: djeck1@jhu.edu)

{\bf Keywords} \\
Jitter, Spike timing, Synchrony, Cross Correlation

\begin{abstract}
  Interval jitter and spike resampling methods are used to analyze the
  time scale on which temporal correlations occur. They allow the
  computation of jitter corrected cross correlograms and the performance of an
  associated statistically robust hypothesis test to decide whether observed
  correlations at a given time scale are significant. Currently used
  Monte Carlo methods approximate the probability distribution of
  coincidences. They require generating $N_{\rm MC}$ simulated spike trains of
  length $T$ and calculating their correlation with another spike
  train up to lag $\tau_{\max}$. This is computationally costly $O(N_{\rm
    MC} \times T \times \tau_{\max})$ and it introduces errors in
  estimating the $p$ value. Instead, we propose to
  compute the distribution in closed form, with a complexity of
  $O(C_{\max} \log(C_{\max}) \tau_{\max})$, where $C_{\max}$ is the
  maximum possible number of coincidences. All results are then exact
  rather than approximate, and as a consequence, the $p$-values
  obtained are the theoretically best possible for the available
  data and test statistic.
 In addition, simulations with realistic parameters predict a
  speed increase over Monte Carlo methods of two orders of magnitude
  for hypothesis testing, and four orders of magnitude for computing
  the full jitter-corrected cross correlogram.
\end{abstract}
\newpage

\section{Introduction}
\label{sec:introduction}

Though there are many means by which neurons communicate, using both
chemical and electrical mechanisms, most attention has been paid to
series of action potentials (spikes). It is known that in some
cases the detailed time structure of these spike trains is used for
information transmission while in others, only the overall number of
spikes in some interval seems to be important but not their position
in the interval. Examples of the first type are various kinds of
``temporal coding'' schemes proposed for different neural system and
for different functional roles, \eg
refs~\cite{Abeles91,Singer_Gray95,Riehle_etal97,Niebur_etal93,Softky95,Steinmetz_etal00a},
while the latter is the well-known rate-code mechanism, \eg
refs.~\cite{Adrian_Zotterman26,Shadlen_Newsome98}. To distinguish
between these two possibilities, it is necessary to find whether
reproducible correlations at the relevant time scale are present in
neuronal data. One common way to approach this problem is to use auto-
or cross-correlation functions as test statistics. Then one can
(a)~search for non-trivial structure in the function, like deviations
from uniformity, and (b)~detect whether there are differences between
these functions in different experimental (\eg behavioral) conditions.

The situation is complicated by the influence of rate variations on
the raw correlations. To increase the signal-to-noise ratio,
correlations are typically computed as averages over many
trials. Changes in the behavioral state of an animal, \eg due to onset
of sensory stimuli or motor responses that occur always at the same
time during a trial, typically result in changes in neural firing
rates which are common to many neurons. While these are genuine
correlations, they are unrelated to the neuronal coding
question. Different techniques have been developed to remove them, \eg
subtraction of a ``shuffle predictor'' \cite{Perkel_etal67b}, the
average of cross correlations between spike trains from permuted
trials\footnote{A ``shift predictor'' is very similar but the
  correlation function is computed from trials that immediately follow
  each other, rather than from randomly selected trials.}.  While this
correction removes correlations that are time-locked to trial onset,
it was later pointed out that peaks in the correlation function that
may be taken as indicative of correlated firing (\eg at zero lag) can
also be caused by slow rate covariations \cite{Brody98c,Brody99}.
After finding a significant peak in the cross correlation function,
this ambiguity can be addressed by analyzing the time scale at which
the measured correlation arises.

It was pointed out more recently~\cite{Amarasingham_etal12} that the
null hypothesis of spike trains being independent in earlier
work~\cite{Perkel_etal67b} is useless if their mean rates
co-vary. Rate co-variation of means the spike trains are not
independent which leads to its immediate rejection of the null without
providing any further insight. Amarasingham and his colleagues instead
proposed a more detailed null hypothesis, namely that within an
interval of width $\Delta$, the exact location of spikes does not
matter. Then, under the null hypothesis, simulated spike trains can be
generated by modifying the spike times of an original measured spike
train within a range of $\Delta$. The cross correlations obtained from
these modified (``jittered'') spike trains are then compared to those
obtained from the original. If significant differences are found, the
null hypothesis is rejected and it is likely that non-random
correlations at time scales $\le \Delta$ are present in the data. Additionally,
this method gives rise to the computation of jitter-corrected cross
correlograms, which have been used to compare changes in synchrony
across experimental conditions
\citep{Hirabayashi_etal13a,Hirabayashi_etal13b,Smith_etal13}.
Because the method relies on repeated simulation of spike trains, it
will be referred to as the {\em Monte Carlo jitter method} for the
purposes of this paper.

While the Monte Carlo jitter method (described fully in Section~\ref{sec:jitter}) is useful and easily generalized
to complex statistical tests and hypotheses, its practical
utility is limited by the inherent trade-off between accuracy and
computation time in all Monte Carlo methods. As we will show in
Section~\ref{sec:simSpeed}, the computation time may be prohibitively
long, and even at this cost, it will only be a numerical approximation
of the true solution. In the case where the test statistic is the  
cross-correllation value at a single lag, the $p$ value can be computed exactly, as was shown 
by \citet{Harrison13}. In the present study, we therefore explore the benefits 
of computing in closed form the distribution which is only approximated by the 
Monte Carlo simulations. Accordingly we refer to this method, described in 
Section~\ref{sec:exact}, as the {\em closed form jitter method}. In addition to 
computing the $p$ value for rejecting the null hypothesis exactly we show that
the computation of the jitter-corrected cross correlogram follows
readily from that derivation. The computational performance of the
closed form jitter and Monte Carlo jitter methods are compared
theoretically (as computational complexity) in
Section~\ref{sec:complexity} and practically (as computational time)
in Section~\ref{sec:simSpeed}.

\begin{table}
\begin{tabular}{l l}
$T$ & Length of binned spike trains \\
$X,Y$ & Two spike trains \\
$\tau$ & Correlation Lag \\
$C(\tau)$ & Correlation of $X$ and $Y$ \\
$\Delta$ & Jitter interval width \\
$i$ & Index over Monte Carlo simulated signals \\
$j$ & Index over jitter intervals \\
$N(X,j)$ & Number of spikes in $X$ in interval $j$ \\
$X^{\rm MC}_i$ & $i$th Monte Carlo simulated signal \\
$N_{\rm MC}$ & Number of Monte Carlo simulated signals \\
$C_i^{\rm MC}$ & Correlation of $X^{\rm MC}_i$ and $Y$ \\
$R_{\tau}$ & Number of cases where $C_i^{\rm MC}(\tau) > C(\tau)$ \\
$p_{\tau}$ & $p$ Value for correlation at lag $\tau$ \\
$\rm JCCG (\tau)$ & Jitter-corrected Cross Correlogram\\
$P_{\tau}(C^{\rm MC})$ & Monte Carlo estimate of the distribution of correlations at lag $\tau$ \\
$C^{\rm int}_j$ & Number of coincidences in the $j$th interval \\
$P_{\tau}(C(\tau))$ & True distribution of correlations at lag $\tau$ \\
$\tau_{\max}$ & Maximum $\tau$ value processed \\
$N_{\max}$ & Maximum value of $N(X,j)$ or $N(Y,j)$ \\
$C_{\max}$ & Maximum possible number of coincidences
\end{tabular}
\caption{{\bf Glossary.} Variables are listed in the order in which they are
introduced.}
\end{table}

\section{Materials and Methods}
\subsection{The Monte Carlo Jitter Method}
\label{sec:jitter}
Utilizing the Monte Carlo jitter method \citep{Amarasingham_etal12}, it
is possible to determine whether correlations arise from fine temporal
structure or larger scale variations, sometimes referred to as rate
covariations. This determination is made by comparing a test statistic (in this 
case cross correlations) of an original pair of spike trains against those 
computed from a set of jittered spike trains as described below.
The jitter method, like cross correlation, operates on
binned spike trains which we take as binary signals with values 0 and
1 and integer arguments $0$ to $T-1$, where $T$ is the
number of bins in the spike train. The binary assumption implies that the bin
size is small enough (typically 1ms or so) such that two spikes cannot
be recorded in a single time bin. A sufficiently small bin size can
always be chosen since there are limits on the minimal inter-spike
interval time due to the refractory period of the neurons in question.

Let $X(t)$ and $Y(t)$ be two such binned spike trains. The processing then 
consists of the following steps:

\begin{enumerate}
\item Compute the cross correlation $C(\tau)$ between the original $X$ 
and $Y$, $$ C(\tau) = \sum_{t} X(t-\tau)Y(t) $$
where the sum runs 
from $0$ to $T-1$ and X is assumed to be $0$ if its argument is outside that range. 
\item Subdivide one of the signals, say $X$, into intervals of width $\Delta$.
\item Count the number of spikes in each interval of $X$. In interval 
 $j$ this is,
\begin{equation}
N(X,j)=\sum_{k = j\Delta}^{(j+1)\Delta-1} X(k)
\label{eq:spikeCount}
\end{equation}
\item For $X$ generate $N_{\rm MC}$ Monte Carlo simulated signals $ \{
  X^{\rm MC}_i \} $, in which the spike counts for each interval are the same
  as in the corresponding interval in $X$, such that $N(X^{\rm MC}_i,j) =
  N(X,j)$ for all $i,j.$ However, now spike times within the
  interval are all equally likely. Spike times should be sampled without
 replacement to ensure that the spike count stays constant without putting multiple spikes in a single bin.
\label{li:genMC}
\item Compute the cross correlation $C^{\rm MC}_i(\tau)$ for lag time $\tau$
  between each $X^{\rm MC}_i$ and the second spike train $Y$ to get an estimate
  of the distribution $P_{\tau}(C^{\rm MC})$ of cross correlation values for
  each time lag $\tau$. $$ C^{\rm MC}_i(\tau) = \sum_{t} X^{\rm MC}_i(t-\tau)Y(t) $$
\item Let $R_{\tau}$ be the number of simulations where $C^{\rm MC}_i(\tau) \geq C(\tau)$. Then the $p$ value for a given lag $\tau$ is computed as $$ p_{\tau} =  \frac{R_{\tau}+1}{ N_{\rm MC}+1}$$
\label{li:PValueMC}
\item If desired, a jitter-corrected cross correlogram (JCCG), defined using 
the expectation operator $E[\cdot]$, can be computed as
\begin{equation}
\rm JCCG (\tau) = C(\tau)-E[C^\mathrm{MC}(\tau)] \approx
C(\tau)-\frac{1}{N_{\rm MC}}\sum_i C^{\rm MC}_i(\tau)
\label{eq:JitterCorrectedGramMC}
\end{equation}
\label{li:JitterCorrectedXcorr}
\end{enumerate}
where the approximation approaches equality with $N_{MC}\rightarrow \infty.$

Jittering the spikes within an interval of size $\Delta$ destroys all
correlations at time scales within this interval. The cross correlations 
computed from the jittered spike trains therefore are not correlated on time
scales $\Delta$ or smaller, and $P_{\tau}(C^{\rm MC})$ is the distribution of
correlations at time lag $\tau$ obtained under the null hypothesis
that correlations at time scales $\le \Delta$ are indistinguishable from
random correlations. If the measured cross correlation $C(\tau)$ is 
significantly outside this distribution, we have to reject the null hypothesis 
and we conclude that nonrandom correlations at lag $\tau$ with time scales 
$\le \Delta$ are found in the observed spike trains. If, on the 
other hand, the observed correlation is consistent with what is seen in the
distribution of jittered spike trains, then we cannot reject the null
hypothesis. This means we cannot exclude that the observed synchrony
is caused by correlations on time scales {\em outside} the jittered
range, in other words that the observed correlation at lag time $\tau$
is caused by rate variations on time scales greater than $\Delta$. 

In practice, $X(t)$ and $Y(t)$ do not have to be gathered from a
continuous block of time. In the case of multiple trials of the same
experimental condition, it may be useful to concatenate the recorded
spike trains (possibly after removing sections of them, like those
recorded during stimulus onsets). In doing so, a period of no spiking
of width $\tau_{\max}$ (the largest correlation lag of interest)
should be added between the trials so that correlations between trials
don't affect the outcome of the jitter procedure.

As mentioned in Section~\ref{sec:introduction}, the practical utility
of the Monte Carlo method is limited by the  trade-off between
accuracy and computation time inherent in all Monte Carlo algorithms. 
Furthermore, in practice a single set of Monte Carlo simulations is often 
generated for many hypothesis tests (\ie tests at multiple lags), introducing
potential dependencies between the different tests when they should be
treated independently \footnote{The procedure of generating one set of spike trains for multiple lags is appropriate inappropriate only if each lag is being tested independently. If the test statistic is the sum of $C(\tau)$ over a range of lag values, a single set of simulated spike trains is appropriate.f}. In order to avoid both of these issues, the
probability distribution $P_{\tau}(C^{\rm MC})$ can be computed
exactly and independently for each time lag as described in the
following.

\subsection{Closed Form Computation}
\label{sec:exact}
\subsubsection{Probability Distribution For One Interval}
\label{sec:algorithm}

First, let us consider a single interval consisting of $\Delta$ time
bins. For example, if spike times have been binned to 2~ms, for 
an interval of width 20~ms we have $\Delta = 10$. 
Since time has been discretized, it is still possible to discuss this unitless value as a length of time, a time scale, or a interval width for a given bin size.
As before, we
assume that the sequence is binary, so each bin has either zero or one
spike. This is true even when the spike times are jittered because
spike times are sampled without replacement.
For this single interval, the probability of a given number of coincidences
occurring is determined by three values: $\Delta$, $N(X,j)$ the number of
spikes in interval $j$  of spike train $X$, and $N(Y,j)$ the number of spikes 
in interval $j$ of spike train $Y.$ 

As a first step, we count the number of perfect coincidences, in which
one spike occurs in both $X$ and $Y$ within the same time bin, meaning
$\tau=0$. Using the standard notation of $a \choose b$ for the
combinatorial operation ($a$ choose $b$), we find that there are $\Delta
\choose N(X,j)$ ways to distribute $N(X,j)$ spikes in $\Delta$ available
bins. The number of empty (spike-less) bins in spike train $Y$ is
$[\Delta-N(Y,j)]$. The number of ways to distribute $N(X,j)$ spikes such
that each of them falls into one of these empty bins is $\Delta-N(Y,j)
\choose N(X,j)$. These are all possible cases in which a coincidence
is avoided. The probability that zero coincidences occur in the $j$-th
interval is therefore
\begin{equation}
  P(C^{\rm int}_j=0|\Delta,N(X,j),N(Y,j))=
  \frac{ {\Delta-N(Y,j) \choose N(X,j)} }{ {\Delta \choose N(X,j)} }
\label{eq:distributionCzero}
\end{equation}
where $C^{\rm int}_j$ is the number of coincidences in this interval. 

We can generalize equation~\ref{eq:distributionCzero} to a non-zero
number $c$ of coincidences by breaking the numerator up into the
number of ways that $c$ spikes can coincide with the spikes in $Y$, and
$N(X,j)-c$ spikes coincide with the gaps (or non-spikes) in $Y$. 
We can thus compute a probability distribution for each
interval $j$,
\begin{equation}
  P(C^{\rm int}_j=c|\Delta,N(X,j),N(Y,j))=
  \frac{ {\Delta-N(Y,j) \choose N(X,j)-c} {N(Y,j) \choose c }}{ {\Delta \choose
      N(X,j)} }
\label{eq:distributionC}
\end{equation}
where we follow the customary convention of setting the value of a
``choose'' operation
to zero if either of its arguments is negative, or if its upper argument
is less than the lower. If this happens in the numerator of
eq.~\ref{eq:distributionC}, the probability on the left
hand side becomes zero. Of course, the denominator is always
positive since $N(X,j)\le \Delta.$ 
This is a hypergeometric distribution.

Equation~\ref{eq:distributionC} is easily generalized to nonzero
values of $\tau$ by applying the analysis leading to it to a shifted
version of $Y$. For the computation of $N(Y,j)$, this implies adding
$\tau$ to the summation limits in eq.~\ref{eq:spikeCount}. As with
other correlation algorithms, the boundaries of finite spike trains
(beginning and end) result in fewer intervals to analyze as $\tau$
gets further away from zero. Thus generalizing eq.~\ref{eq:distributionC}
to non-zero $\tau$, we denote the  resulting
number of coincidences as $ C^{\rm int}_j(\tau)$ and
the associated probability distributions as $P^{\rm int}_{\tau}$.

\subsubsection{Jitter-Corrected Cross Correlation}
\label{sec:exactJCC}

Once we have the analytical probability distribution for the
correlations,
we can obtain all relevant quantities to characterize the pairwise
correlations between two spike trains. It is straightforward to
compute the commonly used jitter-corrected cross correlogram
\cite[e.g.,][]{Hirabayashi_etal13a,Hirabayashi_etal13b,Smith_etal13}
which shows the correlation
function after all correlations on time scales longer than $\Delta$ have
been removed. 
It is defined in analogy to equation~\ref{eq:JitterCorrectedGramMC}
where the expectation value of the stochastic solution,
$E[C^\mathrm{MC}(\tau)]$, was used. We can replace this approximation by the exact
solution $E[C^\mathrm{int}(\tau)]$. Furthermore, by the null
hypothesis each interval is conditionally independent based on the
spike counts. Therefore, the JCCG can be computed
without approximation
 by
\begin{equation}
\rm JCCG (\tau)= 
C(\tau)-E \left[\sum_j C_j^{\rm int}(\tau) \right] = 
C(\tau)-\sum_j E\left[C_j^{\rm int}(\tau)\right]
\label{eq:jitter corrected gram}
\end{equation}
which, as should be remembered, is computed for a specific jitter interval width $\Delta$. The expectation on the right can either be calculated for each window as $N(x,j)\times N(y,j) / \Delta$.

The jitter-corrected cross correlogram is used, for instance, when the
scientific question of interest is whether there are significant
changes in synchrony between conditions, rather than a test of the
presence or absence of synchrony. It is then used as part of a bootstrap
statistical test in which the observed pairwise correlation is
compared with the distribution obtained from eq.~\ref{eq:jitter corrected gram}. 

\subsubsection{Probability Distribution For Spike Train}
One can also obtain the probability distribution for the entire signal
$P_{\tau}(C(\tau))$ as the convolution of the individual probability
distributions for all intervals, $P^{\rm int}_{\tau}$. This is identical to computing
$P_{\tau}(C^{\rm MC})$ from Section~\ref{sec:jitter} with an
infinite number of Monte Carlo simulations for each value of $\tau$.
One can then evaluate how likely it is
that the observed cross correlation $C(\tau)$ is explained by this
probability distribution. The likelihood $p$ that this is the case is
obtained as the integral of the probability density function exceeding
$C(\tau)$, as in 
\begin{equation}
\label{eq:p exact}
p_{\tau}=\sum_{c = C(\tau)}^\infty P_{\tau}(c)
\end{equation}

\section{Results}
\subsection{Computational Complexity}
\label{sec:complexity}

In many situations, the statistical distributions underlying the
phenomena under study are complicated or unknown and performing Monte
Carlo simulations are the only way to make progress, even though it
may be costly and it introduces additional randomness in the
processing. In the case considered here (binary spike trains, null
hypothesis of uniform spike time distribution in fixed interval, cross 
correlation test statistic), the
distribution $P_{\tau}(C)$ can be computed directly, using the
closed form jitter method described above, without the need for
repeated simulations. This section will compare the computational
complexity of using the Monte Carlo jitter method against the direct
computation using the closed form jitter method.

\subsubsection{Monte Carlo Method}
In the Monte Carlo algorithm, the data generation step requires a
permutation of $\Delta$ data points for each interval and simulation. Since a
single permutation operation has a computational complexity $O(\Delta)$,
and $\Delta$ times the number of intervals is the length of the signal
$T$, generating the set of Monte Carlo simulations $\{X_i^{\rm MC}\}$ is
$O(N_{\rm MC} \times T)$. The complexity of cross correlation
or convolution of two signals with lengths $T$ is $O(T\times \log
T)$, assuming  an FFT-based method~\citep{Cooley_Tukey65} is used. 
So computing the full Monte Carlo probability distribution for all
values of $\tau$ is $O(N_{\rm MC} \times T \times \log T)$. In many
cases, not all values of $\tau$ are needed. If the correlation is
computed only for the subset of delay values from 0 to $\tau_{\max}$,
the complexity for the Monte Carlo jitter method is
$$O(N_{\rm MC} \times T \times \tau_{\max})$$

Computing the jitter-corrected cross correlogram by this method only
requires one additional sum, with complexity $O(N_{\rm MC} \times
\tau_{\max})$ so the total complexity remains unchanged.

\subsubsection{Closed Form Probability Distribution}
To compute the exact probability distribution with the closed form jitter
method, note that the values of the distribution can be precomputed
based on the maximum values of $N(X,j)$ and $N(Y,j)$ over all $j$;
call this maximum $N_{\max}$. Also, $n$ choose $k$ operations
can be as fast as $O(\min(k,n-k))$ \citep{Manolopoulos02}. Therefore,
a three dimensional table of all possible values of $P(C^{\rm int}_j| N(X,j),
N(Y,j))$ can be precomputed and then looked up for each
interval. Generating this table requires up to $N_{\max}$ different
values of $C$, $N_{\max}$ values of $N(X,j)$, and $N_{\max}$ values of
$N(Y,j)$. Computing each value requires three $n$ choose $k$
operations, which are on the order of $O(N_{\max})$, so the total computation of the probability table is $O({N_{\max}}^4)$. 
While the exponent is high, the expression does not
have any dependence on the length of the signal and, furthermore,
$N_{\max}\le \Delta$ is a small number in essentially all cases of
interest.
In practice, for analyzing neurophysiological data it is rare that a
time resolution finer than 1~ms is needed, or controlling for cross correlations
at time scales larger than approximately $100$~ms (\ie $\Delta \approx 100$).
 The full lookup
table is therefore maximally a $100\times 100 \times 100$ matrix,
which requires negligible resources to compute and
store.

To compute the combined probability distribution $P_{\tau}(C)$ over all
intervals, all interval probability distributions $P_{\tau}(C^{\rm int}_j)$ must be
convolved, and the computational complexity of the problem is dominated
by these convolution operations. 
As will be shown, we can improve performance by taking advantage of
the structure of the problem at hand, since many of the convolution operations 
are identical. As a result, the convolutions can be grouped together based on 
$N(X,j)$ and $N(Y,j)$ and quickly combined so that $O(T)$ convolutions 
will turn into $O({N_{\max}}^2)$ convolutions. This can be done by the 
following procedure:

\begin{enumerate}
\item  Take the Fast Fourier Transform (FFT) of $P_{\tau}(C^{\rm int}_j| N(X,j), N(Y,j))$ for each 
encountered value of $N(X,j)$ and $N(Y,j)$. 
\label{li:exactFFT}
\item Raise each complex frequency spectrum value to a power equal to the 
number of times that the $(N(X,j), N(Y,j))$ pair appears.
\label{li:exactExponent}
\item  Multiply these frequency spectra.
\label{li:exactCombine}
\item  Take the inverse FFT of the result to get the final probability distribution $P_{\tau}(C)$ and compute $p_{\tau}$ as in equation~\ref{eq:p exact}. 
\label{li:exactIFFT}
\item Repeat steps~\ref{li:exactExponent} through \ref{li:exactIFFT} for each 
value of $\tau$ to be tested.
\label{li:exactRepeat}
\end{enumerate}

The FFT operations in step~\ref{li:exactFFT} must be zero-padded up to
the maximum number of coincident spikes $C_{\max}$ to account for the highest
possible number of synchronous spikes in the combined probability
distribution. Therefore the FFT operation in step~\ref{li:exactFFT} is $O(C_{\max} \times \log(C_{\max}))$.
 In step~\ref{li:exactExponent}, exponentiation is
$O(1)$. However there are $O(T\times {N_{\max}}^2)$ exponents
to be taken, repeated $\tau_{\max}$ times in
step~\ref{li:exactRepeat}. Step \ref{li:exactCombine} requires
$O(T\times {N_{\max}}^2)$ multiplications, again repeated
$\tau_{\max}$ times.  In step~\ref{li:exactIFFT}, the length of the
spectral signal (to be inverted by FFT) is $C_{\max}$,
so the operation is $O(C_{\max} \times \log (C_{\max}))$ repeated
$\tau_{\max}$ times.  When combining these steps, note that $C_{\max}$ is proportional to $T$. However $C_{\max}$ will be used when relevant because it captures the frequency dependence of the computation time. Therefore the total computational complexity is

$$O(C_{\max}\times \log (C_{\max})\times \tau_{\max})$$.

Note that because the zero-frequency component of a probability distribution is 
always exactly unity, the inverse FFT computation will have accuracy limited by 
the precision of the numerical system. In practice this implies that $p$ values 
less than $10^{-13}$ will not be estimated accurately.

\subsubsection{Jitter-Corrected Cross Correlation}
Both the complexity analysis and the actual computation of the jitter-corrected
cross correlogram is much simpler than that of the probability
distribution. We generate a 
lookup table of possible $E\left[C^{\rm int}\right]$ values
and, from equation~\ref{eq:jitter corrected gram}, the
jitter-corrected correlogram can be computed at a speed
of $$O(T \times \tau_{\max})$$.

\subsection{Computational Execution Time}
\label{sec:simSpeed}

For practical applications, consumption of resources is an important
limitation for any computational method. For the size of problems
encountered in typical neurophysiological experiments, the only
limiting resource is execution time.  To compare the performance of
the Monte Carlo jitter and the closed form method, the two algorithms were
run side by side in the MATLAB environment (Mathworks, Natick
MA). Synthetic spike trains were generated that varied in both
frequency of spiking (5 to 500~Hz) and length (1 to 91 seconds). For
each (time, frequency) condition, 50 spike trains were generated,
binned to 1~ms, and the average processing time was computed.
Processing was performed with $\tau_{\max} = 100ms$ and $\Delta = 20$. All
computations were performed on an Intel i7 920 processor with 12~GB of
RAM running Linux Ubuntu 12.04.

For the Monte-Carlo method, $N_{\rm MC}$ was set to 1000. This
selection of $N_{\rm MC}$ is unrealistically low for two reasons. First, it can 
at best result in a Bonferroni corrected $p$ value of 0.201 due to the 201
$p$ values being tested in the range of $-\tau_{\max}$ to $\tau_{\max}$. As
the execution for $N_{\rm MC}=1000$ already takes 5.7 days to
run, increasing $N_{\rm MC}$ is impractical. Second, only a single set of 
Monte-Carlo trials were generated for all lag values computed, inducing 
potential correlations between the $p$ values. These correlations should 
decrease as more trials are generated. Therefore results are
extrapolated to $N_{\rm MC} = 20,000$ (resulting in a minimum $p
\approx 0.01$) under the assumption that the processing for 20
times as many simulations would take 20 times as long. Though the bonferroni 
correction used here is conservative, it is less conservative (by an order of 
magnitude) than simulating a whole new set of spike trains for each $p$ value 
as would be required to entirely eliminate any correlations between the $p$ values.

For the closed form jitter method, all lookup tables were computed {\em de
  novo} for each spike train. This is a conservative approach
(favoring the Monte Carlo technique) since performance of the closed form
jitter method could be improved by computing the tables
only once and using them for all spike trains. This is certainly
advised in a ``production environment.''

\begin{figure}[h]
\centering
\includegraphics[width=1\linewidth]{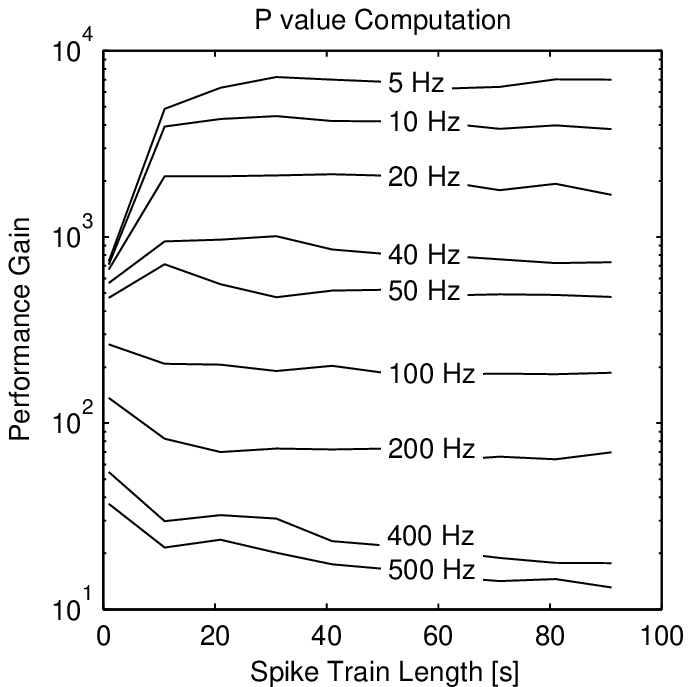}
\caption[Computation time]{Performance gain in implementing the closed form jitter method for $p$ value computations. Gain is defined as the ratio in computation time between the Monte Carlo Jitter method and the closed form jitter method. Processing parameters used are $\tau_{\max} = 100ms$, $\Delta = 20$, and $N_{\rm MC} = 20,000$ (see text for details).}
\label{fig:computation_time}
\end{figure}

The results of this simulation, shown in
Figure~\ref{fig:computation_time}, illustrate a number of features
about the speed of the two algorithms. Plotted is the performance
gain, defined as the ratio of the computation time between the Monte
Carlo jitter method and the closed form jitter method. The first observation
is that the closed form method is substantially faster than the Monte Carlo
method in all cases considered. Second, while the performance gain
depends only weakly on spike train length, it does decrease with
increasing firing rate.
 This is because the computation time of the
closed form jitter method increases with firing rate. 
In practice, however,
it is rare to observe firing at sustained frequencies exceeding 100~Hz
in physiological recordings. In the physiological range, the closed form jitter
algorithm is faster by a factor of approximately 180 to 7200.

\citet{Harrison13} uses importance sampling to accelerate the Monte
Carlo hypothesis testing process which requires drawing fewer
samples. In that work the number of samples needed, even for a low
Bonferroni corrected $p$ value, is reduced to 100. 
However, each sample is reported to take 18 times as long to generate
and process as 
before, effectively resulting in a simulation about 11 times faster than the Monte 
Carlo simulation with $N_{\rm MC} = 20,000$. Therefore, under physiological 
conditions the closed form computation has an expected speed-up of 16 to 650 
times compared to the importance sampling method. It should be noted that in 
cases where even lower $p$ values are needed because of multiple hypothesis 
constraints, importance sampling will provide larger gains in estimating very 
small $p$ values. In these cases, increasing the $p$ value requirements has no 
effect on the computation speed of the closed form method, so the closed form 
method can be expected to be faster in all cases.

\begin{figure}[h]
\centering
\includegraphics[width=1\linewidth]{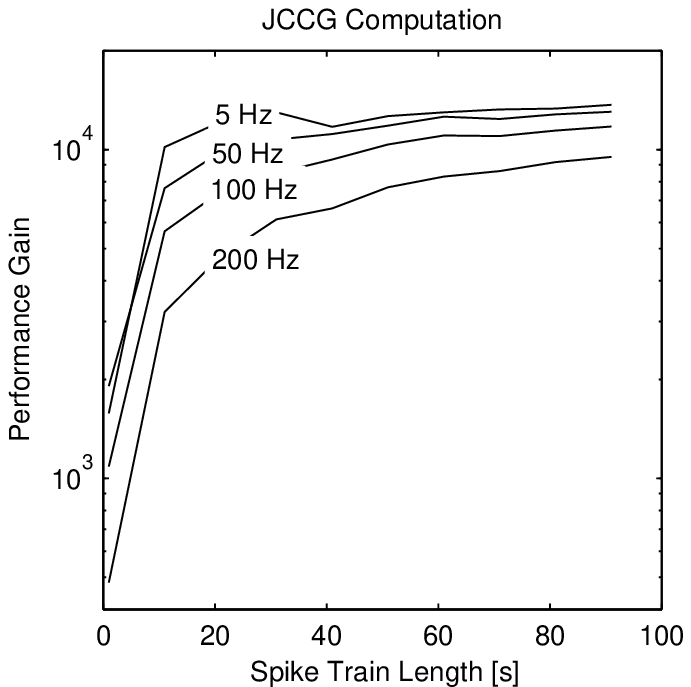}
\caption[Jitter Corrected Correlogram Computation time]{Performance
  gain in implementing the closed form jitter method for Jitter
  Corrected Correlogram computations. Gain is defined as the ratio in
  computation time between the Monte Carlo Jitter method and the
  closed form jitter method. Processing parameters used are
  $\tau_{\max} = 100ms$, $\Delta = 20$, and $N_{\rm MC} = 20,000$ (see
  text for details). The lowered performance gain at signal length of
  1 second is due to the overhead of computing the probability table
  {\em de novo} for each spike train.}
\label{fig:computation_time2}
\end{figure}

Another improvement mentioned in Section~\ref{sec:exactJCC} is the
ability of the closed form jitter method to compute the jitter
corrected correlogram very rapidly, without computing the null
hypothesis distribution of correlation values. To show the magnitude
of the improvement, the simulation was repeated with only the mean of
the null hypothesis distribution computed under the closed form jitter
method since this is all that is needed for the corrected correlation
function, eq.~\ref{eq:jitter corrected gram}. We also restricted firing frequencies to the range
5--200~Hz. Figure~\ref{fig:computation_time2} shows the ratio of the
time it takes to compute equation~\ref{eq:JitterCorrectedGramMC} \vs
equation~\ref{eq:jitter corrected gram}.  In these cases, the
closed form jitter calculation is substantially faster
(480x--13,000x), with increasing benefits for increasing spike train
lengths. As discussed previously, the spike train length is typically
not that of individual trials but of the concatenation of many trials.

\section{Discussion}
\label{sec:Discussion}

The importance, or absence of it, of precise timing of neural spikes
has been discussed for the last half-century. Several techniques have
been developed to characterize neuronal responses at fine time scales
and it is clear that statistical methods have to be developed with
much care to avoid wrong conclusions
\citep[e.g.][]{Gawne_Richmond93,Roy_etal00a}. One important difficulty is that
firing rates can co-vary in the neurons under study. It is well-known
that such co-variations are observable in quantities like pairwise
cross correlation functions but they are typically considered as
irrelevant from the point of view of neuronal coding or of determining
the connectivity in the underlying circuitry. For instance,
the onset of a stimulus will typically generate a temporary increase
in firing rates in sensory cortex but the resulting increase in cross
correlation is usually not considered of importance for neural coding
\cite[for an exception see][who showed that spike timing relative to
onset-related population activity is informative]{Chase_Young07}. One
common way to subtract such stimulus-locked effects is by subtracting
a ``shuffle predictor'' \citep{Perkel_etal67b}, obtained by computing
cross correlations between spike trains from permuted trials. It has
been pointed out repeatedly (see references in the Introduction) that
this does not eliminate spurious correlations, including close to
$\tau=0$ (synchrony).

\citet{Brody98c,Brody99} and \citet{Amarasingham_etal12} proved that
adopting the null hypothesis of independent neurons can not solve the
problem. Observation of such correlations is, indeed, evidence against
the null hypothesis of independence between the two observed spike
trains. Rejection of this null hypothesis can, however, occur either
because spikes in the two spike trains are correlated ``one-by-one''
(synchrony), or because of slow firing rate covariations common to
both spike trains. The fact that this null hypothesis can be rejected does
not tell us {\em why} it is rejected.  If the question is
whether synchrony exists at less than a given time scale (only), this
is the wrong null hypothesis. Instead, the time scale needs to be
specified explicitly. The null hypothesis chosen by
\citet{Amarasingham_etal12} is that changes of spike times within a
time interval of size $\Delta$
have no effect on the computed statistic, in this case the correlation
function. It is this null hypothesis that is tested by computer
simulation in the \citet{Amarasingham_etal12} study and analytically
in this report.

A key element of the methods discussed here is that the jitter
intervals are defined without reference to the original spike
trains. This ensures that if 
the null hypothesis is true, there is no way to distinguish the
original spike trains from the Monte Carlo simulated spike
trains. This characteristic (called exchangeability) ensures that the
obtained $p$ values are from a well formulated hypothesis test. If, on
the other hand, the resampling method was changed so that each spike
was jittered about it's original spike time, then even under the null
hypothesis the original spike train would stand out from the rest
because all of its spikes would be at the center of the jitter
intervals. Therefore the resulting test would not be a proper
statistical test and should be avoided \citep{Amarasingham_etal12}.

We have discussed two ways one can choose to characterize the
correlations between two spike trains. One is a strict hypothesis
testing approach. A null hypothesis is formulated, namely that the
observed correlations are indistinguishable from correlations between
spike trains whose spikes have been distributed randomly within
intervals of length $\Delta$, without changing the number of spikes in each
interval. By comparing the observed correlation with those in the
distribution generated under the null hypothesis, it is then decided
for a given $\alpha$ whether the null hypothesis can be rejected.

The alternative is to compute the time-resolved correlation function
and ``correct'' for the correlations as observed under the null
hypothesis, by subtracting the expectation value of the latter. This
is the more commonly chosen approach, perhaps because the
time-resolved correlation function is both intuitive and
familiar. The distribution of JCCG values can be compared between experimental 
conditions (indicating a change in 'excess synchrony') using a bootstrap test 
to test for significance. Also, its shape (\eg the location of peaks) may 
provide
insight that goes beyond the yes-no answer whether the null hypothesis
can be rejected or not. 

In the \citet{Amarasingham_etal12} study, the Monte Carlo procedure is
further developed to account for more potential causes of fine timing
effects besides synchrony such as ramping spike rates within an
interval or inter-spike interval distribution effects.  
These methods are straightforward and statistically well-defined. Like
any Monte Carlo method, however, they only generate an approximation
to the underlying distribution whose quality depends on the number of
surrogate spike trains. In practice what is more problematic is that
the method can be computationally very costly. For instance, as
discussed in section~\ref{sec:simSpeed}, our example problem using the
simplest of the null hypotheses discussed (50 spike trains of a few
seconds long each, mean rates between 1 and 100~Hz, maximal time lag
of 100~ms, $\alpha=0.01$ with Bonferroni correction applied) would
have required a simulation several {\em months} long on a reasonably
fast machine. We therefore only simulated $N_{MC}=1000$ trials and
extrapolated to the execution time needed for $N_{MC}=20,000$ but even
that abbreviated Monte Carlo run took nearly six days. Some progress
can be made by using much faster machines or many machines (the
problem parallelizes easily) but execution time is clearly a problem.

In contrast, the closed form jitter methods this report focuses on are
exact, rather than approximate. More important for practical
applications may be that they are extremely efficient, with a speed-up
of at least two orders of magnitude for the hypothesis testing
approach, and four orders of magnitude for the full correlation
functions. Even over importance sampling methods \citep{Harrison13},
they have been shown to provide a substantial increase in speed. For
the hypothesis testing examples used in our study (whose scope is
quite comparable to that of typical neurophysiological experiments,
assuming a proper Bonferroni correction is applied), computation time
is reduced from more than 100 days under the original Monte Carlo
method to about one night. Computational time required for the full
correlation function is reduced from over 100 days to a few
minutes. An increase in performance on this scale is more than merely
a quantitative improvement. For instance, it is essentially impossible
to explore variations in the analyses (like the influence of the
jitter time scale $\Delta$) if each computational run takes a few
months, but it is easy to do if it takes minutes.

So far we were only concerned with correlations between two spike
trains. Modern recording techniques are already increasing the number
of simultaneously recorded spike trains to tens or
hundreds. Unfortunately, the closed-form jitter method is limited in the 
ability to analyze large ensembles. This is because the correlation functions 
of some pairs in an ensemble will restrict the possible correlation values of 
other pairs. For example, if there are three neurons $X$, $Y$, and $Z$, and the 
pairs $XY$ and $YZ$ have perfect correlation, then the pair $XZ$ must also have 
perfect correlation. A Monte-Carlo jitter analysis that jitters an entire 
ensemble of neurons and then performs a hypothesis test on the ensemble can be 
performed relatively simply, but no such closed-form method exists yet. In order to avoid the constraints of the type described above, only $N-1$ pairs of neurons can be analyzed with closed form methods when $N$ neurons are recorded.

Additionally, the nature of the exact solutions provides an
opportunity for further exact analysis. Having a closed form solution allows
questions about the effects of spike sorting errors, the value of
$\Delta$, or the structure of $JCCG(\tau)$ to be addressed rigorously
and more precisely than is possible with any numerical method.

In conclusion, we study a statistical framework for quantifying
correlations between spike trains at given time scales. It can be
applied both for hypothesis testing and for correcting observed
correlation functions for correlations at these time scales. Results
are exact, and both computational complexity and computational time for
realistic examples are several orders of magnitude lower than related
approaches based on Monte Carlo simulations.

 Matlab code will be made available  by the authors upon request.

\section*{Acknowledgments}
This work was supported by the Office of Naval Research under MURI
grant N000141010278 and by NIH under R01EY016281-02. We acknowledge
discussions with Drs. R\"udiger von der Heydt and Anne Martin who also
gave us access to unpublished data.


\end{document}